\documentclass[osajnl,twocolumn,showpacs,superscriptaddress,10pt]{revtex4-1} 
\usepackage{amsmath,amssymb,graphicx}

\begin{document}

\title{Weak Values Technique for Velocity Measurements}

\author{Gerardo I. Viza}
\email{gerviza@pas.rochester.edu}
\author{Juli\'{a}n Mart\'{i}nez-Rinc\'{o}n}
\email{jrmartir@pas.rochester.edu}
\author{Gregory A. Howland}
\affiliation{Department of Physics and Astronomy, University of Rochester, Rochester, New York 14627, USA}
\author{Hadas Frostig}
\affiliation{Department of Physics of Complex Systems,
Weizmann Institute of Science, Rehovot 76100, Israel}
\author{Itay Shomroni}
\author{Barak Dayan}
\affiliation{Department of Chemical Physics, Weizmann Institute of Science, Rehovot
76100, Israel}
\author{John C. Howell}
\affiliation{Department of Physics and Astronomy, University of Rochester, Rochester, New York 14627, USA}
\affiliation{Department of Physics of Complex Systems,
Weizmann Institute of Science, Rehovot 76100, Israel}

\begin{abstract}In a recent letter, Brunner and Simon propose an interferometric scheme using imaginary 
weak values with a frequency-domain analysis to outperform standard 
interferometry in longitudinal phase shifts~[N. Brunner and C. Simon, Phys. Rev. Lett {\bf105} (2010)]. Here we demonstrate an interferometric scheme combined with a time-domain analysis to measure longitudinal velocities. The technique employs
the near-destructive interference of non-Fourier limited pulses, one Doppler shifted due to a moving mirror, in a
Michelson interferometer. We achieve a velocity
measurement of $400$ fm/s and show our 
estimator to be efficient by reaching its Cram\'{e}r-Rao bound.
\end{abstract}


\maketitle 

\emph{Introduction.}--\:In information theory, the
Cram\'{e}r-Rao bound (CRB)~\cite{Rao,Pfister} is the fundamental limit in the minimum 
uncertainty for parameter estimation. Measurements of
phase~\cite{Starling3,Li,Brunner}, beam
deflection~\cite{HowellTheory,Dixon}, pulse arrival
time~\cite{Giovannetti}, Doppler shift~\cite{Kaczmarek,Gibble} and velocity~\cite{Meier,Charrett,Czarske,Scalise} are all fundamentally bounded by a CRB. 
If a measurement technique reaches the CRB, its estimator is said to be efficient.

Spurred by fundamental studies of quantum phenomena~\cite{Hosten} and
by developments in precision
measurements~\cite{Brunner,Starling2,Starling3,Dixon}, the field of
weak values~\cite{Aharonov,Feizpour,Kocsis,Bamber,Popescu} has become a powerful tool 
for parameter estimation~\cite{Hofmann,Julian,Strubi,Kedem}.
The precision measurements inspired by weak values are not
necessarily new or quantum.  For example
Zernike's phase contrast imaging~\cite{Zernike}, awarded the $1953$
Nobel prize, can be classified as a weak value technique. An important aspect
of the weak-values framework is that it provides a methodology for mitigating technical
noise and amplifying an effect in one domain that is technologically
difficult to observe in the conjugate domain.

Recently, Simon and Brunner showed that a 
weak-values technique allows us to observe a large spectral shift induced by a small temporal shift. They also showed their technique
outperforms standard interferometry limited by technical noise~\cite{Brunner}. Here we consider the opposite
regime, where a small spectral shift causes a
large temporal shift. The spectral shift in our experiment is a Doppler frequency shift produced by a moving mirror.
Using established interferometry to measure velocities arriving at the CRB is difficult 
but achievable as seen in Ref~\cite{Pfister}. Our technique is comparable to 
standard interferometry, but allows us to reach its' CRB with a relatively simple 
method in a regime where $1/f$ noise typically dominates. In this letter 
we show a weak-value optical technique to measure sub pm/s velocities. The protocal 
reaches the predicted CRB, averting technical noise and experimental imperfections.

\begin{figure}[htb]
 \centering
 \includegraphics[scale=0.6]{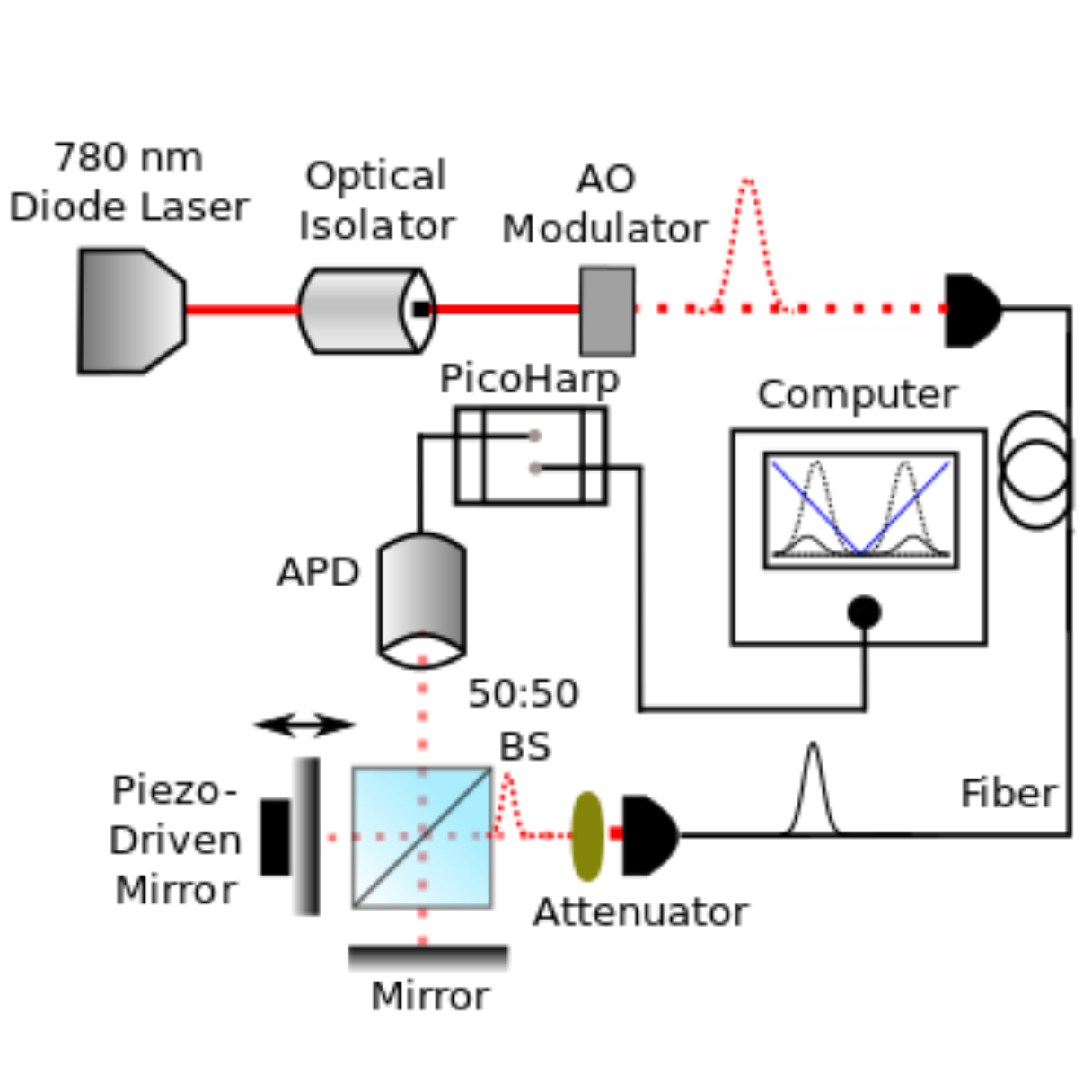}
 \caption{An optical modulator generates a non-Fourier limited Gaussian shaped pulse. We
   couple the pulse to a fiber and launch it to a Michelson
   interferometer where one mirror is
   moving with constant speed $v$.
   The interference is controlled
   by inducing a phase offset $2\phi$ with the piezoelectric
   mirror. Photons exiting the interferometer are coupled into a fiber (not shown) and 
   the arrival time of single photons are measured with an avalanche photo diode (APD) and a photon counting module.}
 \label{fig:setup}
\end{figure}

\emph{Theoretical description.}--\:The protocol, shown in Fig.~\ref{fig:setup}, uses a non-Fourier limited 
Gaussian pulse (i.e., $c\tau\gg$ coherence length of the laser where $\tau$ is the length of the pulse). The pulse, with initial intensity profile
$I_{in}(t)=I_0\exp\left(-t^2/2\tau^2\right)$, is sent through a
Michelson interferometer with a slowly 
moving mirror in one arm.
The interferometer is tuned slightly
off destructive interference by an amount $2\phi$, such that
the output signal takes the form
\begin{eqnarray}\label{eq:Intensity1}
I_{out}(t)&\propto & I_{in}(t)\left|1-\exp\left(i2\phi+i2kx(t)\right)\right|^2\nonumber\\
&\propto & I_0\exp\left({-t^2/2\tau^2}\right)\sin^2\phi\left|\frac{\sin\left(\phi+kvt\right)}{\sin\phi}\right|^2,
\end{eqnarray}
where $k=2\pi/\lambda$, $x(t)=vt$ and $v$ is the velocity of the mirror. Assuming $kv\tau\ll\phi$, making a small angle approximation of $\phi$ and re-exponentiating the output intensity, we obtain
\begin{equation}
 I_{out}(t)\approx \left(I_0\sin^2\phi\right)\exp\left[-\frac{1}{2\tau^2}\left(t-\frac{2kv \tau^2}{\phi}\right)^2\right].
\label{eq:Intensity2}
\end{equation}
Near-destructive interference reduces the peak intensity of the pulse by a factor $\sin^2\phi$, which is the probability for a single photon passing through the interferometer to
reach the detector. Importantly, a time shift in the peak output intensity, $\delta t=2kv\tau^2/\phi$, has been induced with respect to the input. The velocity $v$ can be obtained from measurements of the time shift $\delta t$.

We can rewrite the time shift in Eq.~(\ref{eq:Intensity2}) in terms of the spectral shift $\delta t=2kv\tau^2/\phi=2\pi f_d\tau^2/\phi$ where the spectral shift, $f_d=2v/\lambda$, of the pulse is proportional to velocity $v$. Instead of a direct spectral measurement, we obtain the velocity by measuring the induced time shift of the non-Fourier limited pulses. The time shift is amplified in the measurement of $v$ which is accompanied by a decrease in the measured intensity. These two results are well-known properties of the interferometric weak value amplification technique. In fact, a full weak value description, using coherent Fourier limited pulses, can be formulated obtaining an identical result to Eq.~(\ref{eq:Intensity2}). In our case, the use of non-Fourier limited pulses allows us to produce large time shifts regardless of the laser linewidth.

We now consider the fundamental limitations of our velocity measurement set by the CRB. The CRB is equal to the inverse of the Fisher 
information, the amount of information a random variable (arrival time of photons) provides about a parameter of interest (velocity). Assume that $N$ photons are sent through the interferometer. We want to determine the shift $\delta t$ from the set of $N\sin^2\phi$ independent measurements of photon arrival times. Such measurements follow the distribution $P(t;\delta t)=\left(2\pi\tau^2\right)^{-1/2}\,\exp{\left[-(t-\delta t)^2/2\tau^2\right]}$. The Fisher information is
\begin{equation}
F(\delta t)=N\sin^2\phi\int dt\,P(t;\delta t)\,\left[\frac{d}{d\,\delta t}\ln{P(t;\delta t)}\right]^2\approx\frac{N\phi^2}{\tau^2}.\label{eq:FisherInformation}
\end{equation}
The CRB, $F^{-1}$, is the minimum variance~\cite{book1,book2} of an
unbiased estimation of $\delta t$. The sensitivity in
the determination of $\delta t$ is therefore bounded by $\Delta\left(\delta t\right)\geq\tau/\phi\,\sqrt{N}$. The error in the estimation of $v$ is then bounded by 
\begin{equation}
\Delta v_{CRB}=\frac{\Delta\left(\delta t\right)\,\phi}{2\,k\,\tau^2}  =\frac{1}{2 k \tau\sqrt{N}}. \label{CRB}
\end{equation}
Note that this minimum uncertainty is independent of the actual value of $v$ measured. This also determines the smallest resolvable velocity, when the signal-to-noise ratio is unity. 
The signal-to-noise ratio is
\begin{equation}
\mathcal{SNR} = \frac{\delta t}{\tau}\phi\sqrt{N}=\frac{v}{\Delta v}=\frac{f_d}{\Delta f_d}.\label{eq:SNR}
\end{equation}


\emph{Experiment.}--\:We use a grating feedback laser with $\lambda\approx780$ nm. An acoustic optical
modulator creates Gaussian pulses of length $\tau$ which we couple into a fiber. We launch them
through the $50$:$50$ beam splitter (BS) of the interferometer.
The piezoelectric actuated mirror is driven by a triangle
function with frequency $f_m$ and peak-to-peak voltage
$V_{pp}$. The pulse length is smaller than
half the oscillating mirror period, so that a pulse experiences
a single, constant velocity. An opposite constant velocity is observed for each sequential
pulse because the sign depends on whether the mirror
moves toward the BS, or recedes away (see Fig.~\ref{fig:setup}).
The piezoelectric response $\alpha$ is calibrated by
varying the voltage to change the dark port to a bright port. The piezo response was found to be
$\alpha\approx 27$ pm/mV for a low frequency-voltage product.
The arm lengths (beam splitter-mirror distances)
are approximately $1$ mm (not including the BS size) to ensure long term phase
stability. Photon arrival times are recorded with an avalanche photon diode (APD) and a photon counting module (PicoQuant PicoHarp $300$). The detector collects arrival times with $350$ ps resolution.

To calibrate the experiment we record the number of detected photons entering the interferometer, $N$. Then, the piezo-driven mirror is biased near destructive interference and fed a triangle signal. We calculate the mean and error of the arrival time of the $N_\phi$ detected photons for each set of pulses. The mean of the Gaussian determines the time shift $\delta t$ from which
the velocity is extracted, and the angle $\phi\approx\sqrt{N_\phi/N}$ is calculated. Lastly, to reach the CRB, we attenuate the peak of the pulses to about a million photons a second.

\begin{figure}[h]
  \centering
  \includegraphics[scale=0.4]{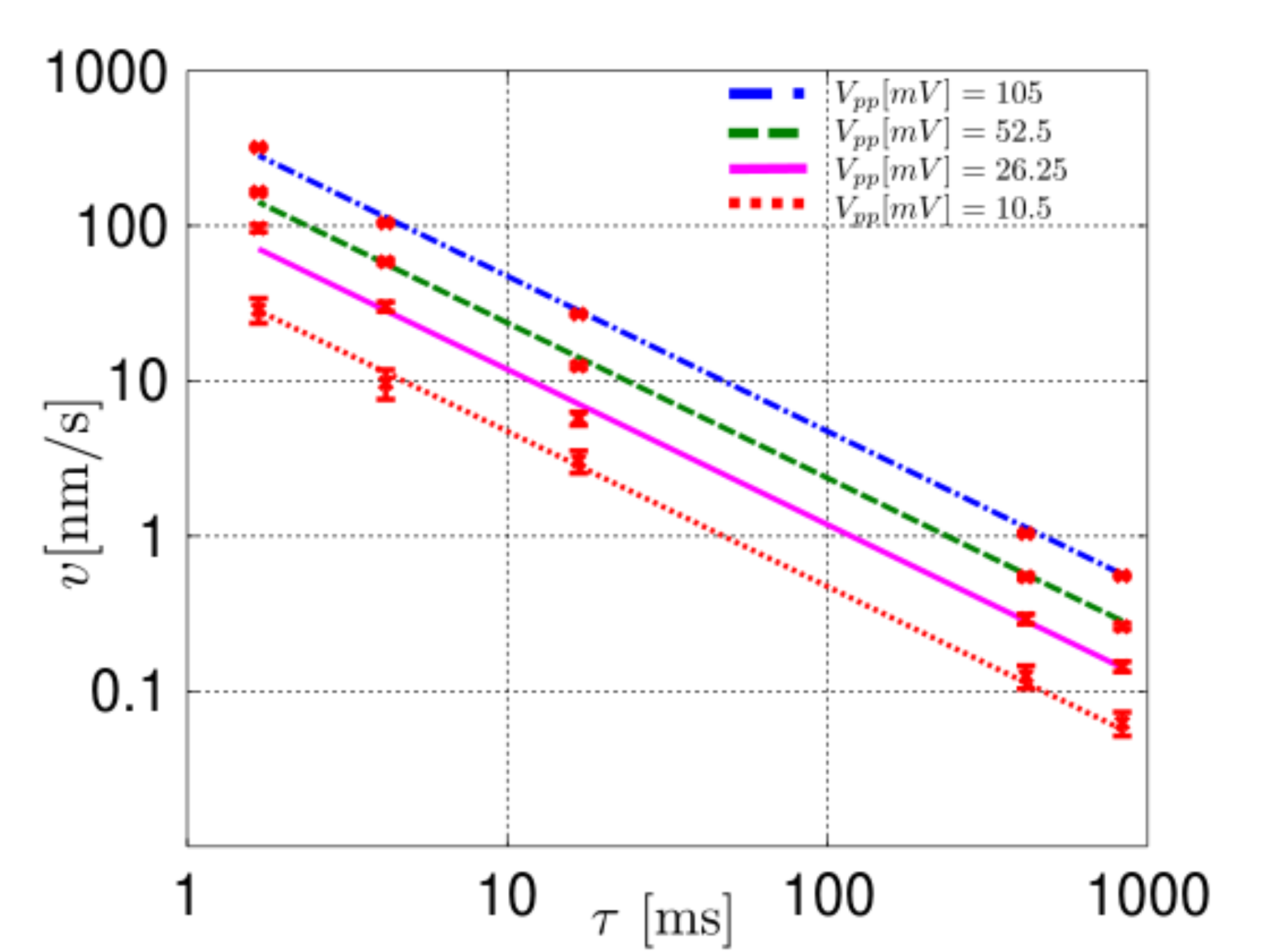}
  \caption{(color online). The Doppler shift, $f_d$, is plotted as a
    function of $\tau=\{1.67$, $4.17$,
    $16.7$, $417$ and $833\}$ ms.
    The phase offset angle is $\phi=0.31$ radians. The points are the experimental results and the lines are the theoretical predictions for different voltages. Signal-to-noise ratios are 54, 27.4, 14.7, and 5.7 for $Vpp=$\{105, 52.5, 26.25, 10.5\} mV respectively.}
  \label{fig:Data1}
\end{figure}

\emph{Results.}--\:We present velocity measurements $v$ as a function of the pulse width $\tau$ for different 
amplitudes on the moving mirror in Fig.~\ref{fig:Data1}. The lines are the theoretical 
predictions, $v=2f_mV_{pp}\alpha$, where $2f_m=1/{6\tau}$. The mirror voltages are $V_{pp}=\{105$, $52.5$, 
$26.25$, $10.5\}$ mV and angle is $\phi=0.31\pm0.02$ rad. 
The results agree well with the theoretical predictions. The smallest measurement of velocity in Fig.~\ref{fig:Data1} is $v=60\pm11$ pm/s. The angle $\phi=0.31$ might seem large; however, comparing the exact form in Eq~(\ref{eq:Intensity1}), $|\sin(\phi + kvt)/\sin(\phi)|$, to the approximation, $|\exp(kvt/\phi)|$, shows a discrepancy less than $1\%$ for the experimental parameters. 

\begin{figure}[htb]
 \centering
 \includegraphics[scale=0.4]{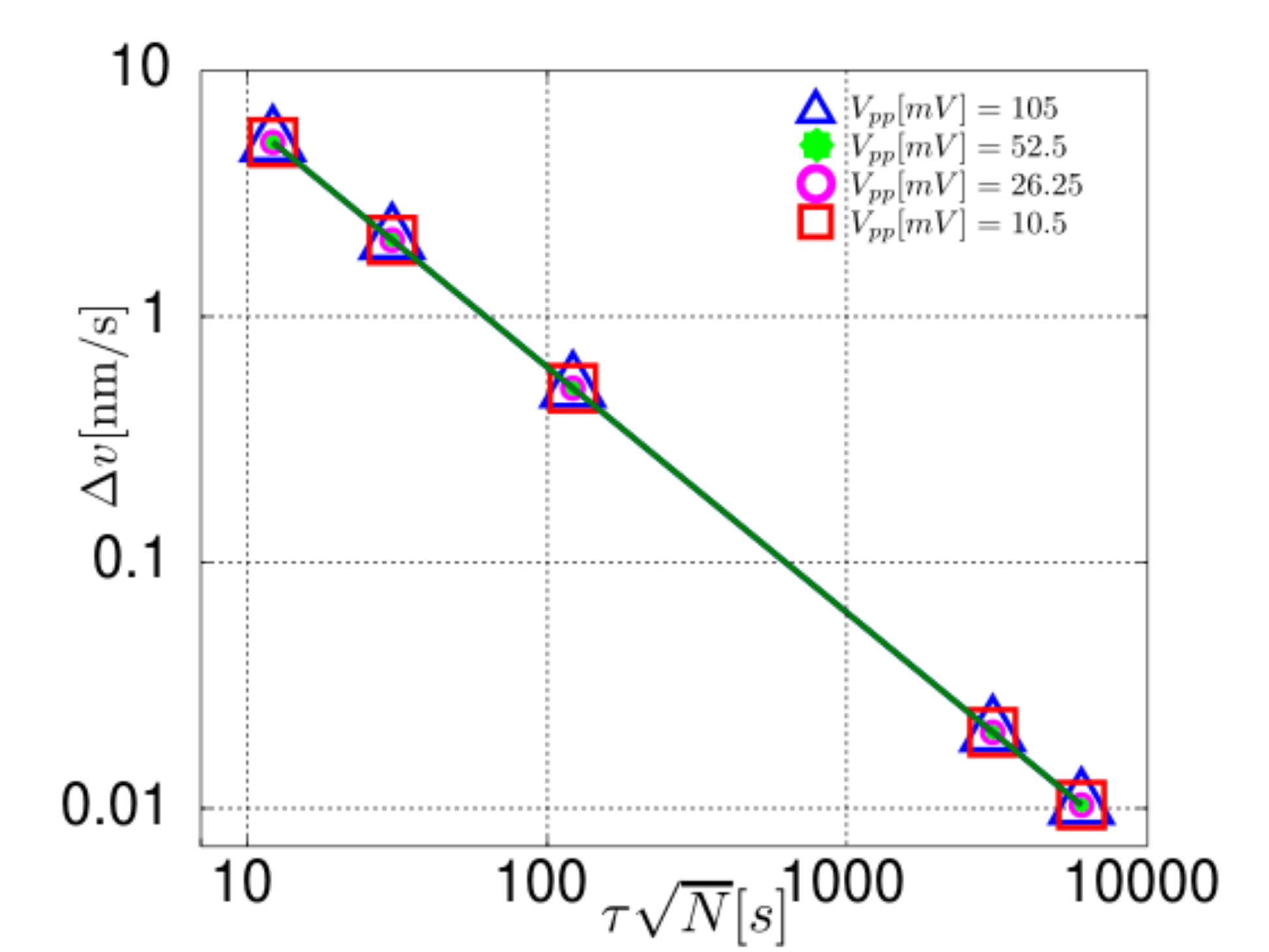}
 \caption{Experimental error in Fig.~\ref{fig:Data1} as a function of $\tau\sqrt{N}$. The solid line is the CRB as in Eq.~(\ref{CRB}) for $N\approx54\times10^6$ photons. Note there are no error bars.}
 \label{fig:Data2}
\end{figure}

The uncertainties of the measurements in Fig.~\ref{fig:Data1} are plotted separately in Fig.~\ref{fig:Data2} and compared to the CRB Eq.~(\ref{CRB}). 
The error matches the CRB, thus the estimator is efficient and no other estimator can produce smaller uncertainties. This technique did not require noise filters or frequency locking to reach the fundamental uncertainty in the mean arrival time of the photons. In addition, the fluctuations in the post selection angle $\phi$  are
negligible. Therefore, our velocity measurement is fundamentally
bounded by its CRB.

It is important to note our CRB is scaled by the
maximum number of detected photons $N$. The collection-detection efficiency is about $20\%$ due to the $50$:$50$ BS (not shown in Fig.\ref{fig:setup}) located before the APD 
used for alignment of the dark port, the efficiencies of the APD and the fiber coupling. Our calculations do not take the 
collection-detection efficiency into account.

The results show precise and accurate detection of velocity measurements in the pm/s
range. Results from Fig.~\ref{fig:Data1} show smaller velocities can be measured with longer pulses.

Now we seek to achieve the smallest velocities without 
the concern of reaching the CRB. Consider the temporal shift, advance or delay, of the
pulse exiting the interferometer. Since the peak of the
pulse is sufficient to detect a the shift, we require a small region around
the peak to determine the shift.
This allows the use of effectively large values of $\tau$ without requiring long term interferometric stability. Since the pulses are non-Fourier transform limited it is not necessary
to use an entire Gaussian pulse. We truncate the Gaussian pulse to a width of $\tau$, that is $2f_m=1/\tau$. In other words, the light intensity into the interferometer never drops below the $88\%$ of the peak intensity and there is $12\%$ peak to peak intensity variation following the peak of the Gaussian profile. 

We show velocities in the sub pm/s range using truncated pulses in Table~\ref{tab:table1}. The mirror frequency was set to $10$ mHz, which corresponds to $\tau=50$ s, and data was taken for voltages peak to peak, $V_{pp}=\{2$, $1$, $0.5\}$ mV, for the piezo driving the mirror. Data was collected in intervals of $10$ minutes (due to drift instability in intensity), and $13$ sets of data were taken for each voltage. We did a Gaussian fit for each $10$ minute interval. The time shift and its error were found as the mean and standard deviation respectfully of the 13 time shifts obtained.
The time shift was in the $10$ s of millisecond and corresponds to small Doppler shifts in the mircoHz range. This leads to the best technical noise limited measurement of $(400 \pm 400)$ fm/s. Nevertheless, both accuracy and precision are lost due to numerically fitting the truncated distributions. Note that the measurements are all relative velocities because of the oscillating mirror. In one period there would be two pulses each with opposite but equal speeds.

The results remain consistent with the full Gaussian picture theory, Eq.~(\ref{eq:Intensity2}), but not with the CRB theory in Eq.~(\ref{CRB}). Calculating
the mean arrival time of the photons is not a good estimator of the time shift because 
we lack the full Gaussian pulse profile. Therefore we numerically fit the data to a unnormalized function $A\exp\left[-(t-\delta t)^2/{2\tau^2}\right]$, the shift $\delta t$ is extracted and the velocity, $v$, is backed out.

\begin{table}[h]
\begin{ruledtabular}
\begin{tabular}{cccc}
\textrm{$V_{pp}$ [mV]}&
\textrm{$\phi$ [$\pm0.002$ rad]}&
\textrm{$f_d$  [$\mu$Hz]}&
\textrm{$v$  [pm/s]}\\
\colrule
$2.0$ & $0.275$ & $3.6\pm1.2$ & $1.4\pm0.5$\\
$1.0$ & $0.276$ & $1.6\pm1.1$ & $0.6\pm0.4$\\
$0.5$ &  $0.279$ & $1\pm1$  & $0.4\pm0.4$\\
\end{tabular}
\end{ruledtabular}
\caption{\label{tab:table1}%
Results of the cut Gaussian profile with $\tau = 50$ s and $N\approx66\times10^9$. The collection-detection efficiency is about $20\%$. The error is from the statistics of numerically fitting each run. Integration time was about two hours worth of data. }
\end{table}
\emph{Conclusion.}--\:In this letter, we show using non-Fourier limited pulses and standard interferometry inspired by weak values, sub pm/s velocities can be measured.
Using a Michelson interferometer tuned near a dark port we measure velocities as low as $400 \pm 400$ fm/s.
We accomplished sub pm/s velocity detection by bypassing the technical noise that flood intensity detectors to reach the CRB. The uncertainty of the phase measurement is negligible when compared to the uncertainty of $v$ for our parameter values so our uncertainty is the fundamental limit. Finally the error
in our measurement of $v$ matches the predicted CRB making this estimator efficient and the ultimate limit in uncertainty for velocity measurements.

This work was supported by the Army Research Office grant number
W911NF-12-1-0263.


\end{document}